\documentclass[twocolumn,twoside,slac_two]{revtex4}
\usepackage{graphicx}
\usepackage{fancyhdr}
\usepackage[utf8]{inputenc}
\usepackage{pstricks}
\usepackage{subfigure}
\usepackage{mathrsfs}
\usepackage{amsmath}
\usepackage{mathbbol}
\usepackage{amsbsy}
\usepackage{amssymb}
\usepackage{units}
\usepackage{fancyhdr}

\newcommand{\mr}{\mathrm}

\begin{document}

\title{Production measurements at LHCb with the first data}

%

\author{F. Dettori (On behalf of the LHCb
Collaboration)}
\affiliation{Università degli Studi di Cagliari and INFN, 
S. P. Monserrato Sestu Km 0.700, 
 09042 Monserrato (CA) Italy
}

\begin{abstract}
We report on the perspective measurements of inclusive particle production in
high-energy p-p collisions with data to be collected by the LHCb experiment at
CERN's LHC. These include V0 and D meson production studies, which can be based
on a minimum bias sample, as well as charmonia production studies, which need a
muon-triggered samples. Using reconstructed $J/\psi \to \mu^+ \mu^-$ decays,
both the prompt $J/\psi$ and $b\to J/\psi$ production cross-sections will be
determined, in the
forward pseudo-rapidity range of 2-5 covered by LHCb. Due to the large
production rate, such analyses will be possible with very small integrated
luminosities of the order of a few $\mr{pb}^{-1}$. 
Other charmonia related measurements
will also be discussed, such as that of the $J/\psi$ polarization at production
or of the production of some of the new X, Y and Z states.
\end{abstract}

\maketitle

\thispagestyle{fancy}

\section{Introduction}

Dedicated to the study of the $b$-flavour quark sector, the LHCb experiment 
will take data in proton-proton collisions at the CERN's LHC.
At an energy in the center of mass of 14 TeV, the cross section for $b \bar b$
pair production is 
$500 ~\mr{pb^{-1}}$, so that, with a nominal luminosity of $2 \cdot 10^{32}
~\mr{cm^{-2}
s^{-1}}$, $10^{12}$ $b \bar b$ pairs will be produced in one year ($10^7$ s) of
data taking.
Within this frame the CP violation and rare decays will be studied and CKM
matrix tests will be performed on the full $b$-hadrons spectrum as well as in
the $charm$ sector in search for hints of new physics \cite{ref:TP,ref:TDR}.
The detector of the LHCb experiment is presented in §\ref{sec:detector}.

It is expected that the 2009 - 2010 LHC data taking will start at the center of
mass energy $\sqrt{s} = 7$ TeV (may go up to 10 TeV) and an integrated
luminosity of 100 to 200 $\mr{pb}^-1$ is anticipated.
With these conditions around $10^8$ minimum bias
events will be collected during the first days of data taking.
These data will allow to perform various production studies on V0 particles,
$D$ mesons and $J/\psi$'s, which will be presented in detail in this article.

\section{LHCb detector \label{sec:detector}}

\begin{figure}[t]
\includegraphics[width = 8 cm]{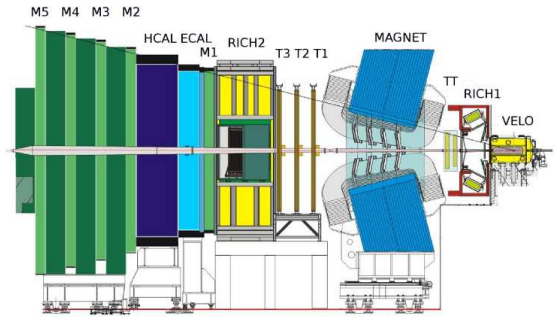}
\caption{Layout of the LHCb detector.}\label{fig:layout}
\end{figure}

The LHCb detector is a single-arm spectrometer placed in the forward region of
the p-p interaction point. With an angular coverage, with respect to the
beam-axis, from 10 (15) to 300 (250) mrad in the bending (non-bending) plane, it
has an acceptance from 1.9 to 4.9 in rapidity, suited to collect the $b \bar b$
quarks production which, at the LHC energies, is well peaked and correlated in
the forward and backward region. Here, just a brief description of the LHCb
detector is given, more details can be found in Ref.~\cite{ref:jinst}.

The LHCb detector is divided in several subdetectors, a simple layout is
shown in Fig.~\ref{fig:layout}. Around the
interaction point is the Vertex Locator, a silicon detector that measures the
radial and angular positions of charged tracks. The momentum measurement is
ensured by the dipole magnet (integrated field of 4 Tm) and the 
tracking system. The latter is subdivided in a Trigger Tracker (TT), a silicon  
micro-strips detector, and three
tracking stations (T1-T3) made up of
silicon micro-strips for the inner part and straw tubes for the outer part.
Particle identification, and in particular $\pi-K$ separation, is ensured by
two Ring Imaging Cherenckov detectors, RICH1 and RICH2. Identification of
muons is given by the MUON system, composed by one detector station (M1) placed
upstream of the calorimeter system and 4 downstream (M2-M5); the stations are
build up from MWPC's with the exception of the very inner part of M1 where
triple-GEM detectors are exploited.
Finally, energy measurement is made by the calorimeter system: a Scintillator
Pad Detector (SPD) and Pre-Shower (PS), the shashlik Electromagnetic
Calorimeter (ECAL) and a hadronic calorimeter (HCAL) with Fe and scintillator
tiles. A summary of the expected performances of LHCb is shown in
Table~\ref{tab:performance}.

\begin{table*}[t]

\caption{Résumé of the LHCb detector performances.}\label{tab:performance}
 \begin{tabular}{llll}
Description 			& Performance \\ 
\hline
Momentum resolution 		& $\sigma(p)/p \sim 0.4 \% $ \\
Energy resolution (ECAL) 	& $\sigma_E / E \simeq 9\%/\sqrt{E} \oplus
0.8\%$  \\
Energy resolution (HCAL) 	& $\sigma_E / E \simeq 69\%/\sqrt{E} \oplus
9\%$\\
$b$-hadrons mass resolution 	&  $\sigma(M) \sim 14 ~\mr{MeV/c^2}$ \\
Primary [secondary] vertex position &  $\sigma(\vec x) \sim 50 [150] ~\mr{\mu
m}$
\\
Impact parameter 		&  $\sigma(IP) \simeq 14 +\nicefrac{35}{p_T
\mr{(GeV/c)}} ~ \mr{\mu m}$\\
Time resolution  on $b$-hadrons proper lifetime &  $\sigma(t) \sim 40
~\mr{fs}$ \\
Kaon identification 		& $\varepsilon(K) \sim 95\%$ at $5\%$ of
$\pi/K$ mis-id.\\
Muon identification 		& $\varepsilon(\mu) \sim 94\%$ at $3\%$ of
$\pi,K/\mu$ mis-id.\\
 \end{tabular}

\end{table*}

A fundamental feature of LHCb is its trigger system. The rate reduction from
40 MHz, LHC bunch crossing frequency, to 2 kHz, at which events are written on
tape for later analysis, is done by two trigger levels. The Level 0
trigger is hardware based, build up of custom made electronics, and reduces the
rate from 40 to 1 MHz mainly requiring particles with high transverse momentum
($p_T$) respect to the beam direction in the calorimeters and muon system. The
High Level Trigger will exploit the full event
data from the detector to select events at 2 kHz. It is software based and its
algorithms will evolve with the knowledge of the apparatus performance and
physics programme leading to high flexibility.

\section{Physics with minimum bias}

\begin{figure}
\begin{center}
 \includegraphics[width = 8 cm ]{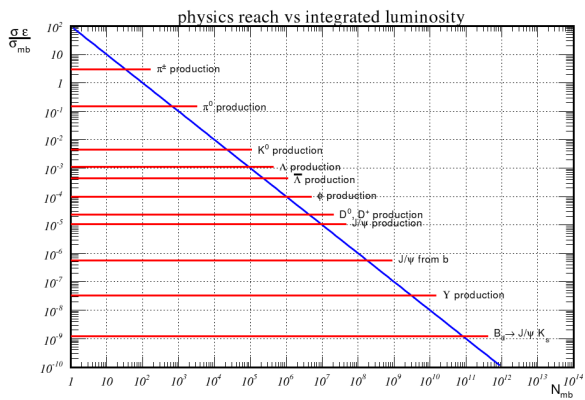}
 \caption{Physics reach at LHCb as a function of the number of
minimum bias events collected. For each channel the relative cross-section
with respect to minimum bias times the efficiency is shown. The straight line
corresponds to 100 events for each channel.
}\label{fig:physicsreach}
\end{center}
\end{figure}

As already said the LHC 2009-2010 run conditions will not be the nominal ones,
nevertheless, as soon as the collisions start, lot of interesting physics will
be available to be studied. Already within the minimum bias events, many
measurements can be done: in
particular in Fig.~\ref{fig:physicsreach} is shown the physics reach as a
function of the number of minimum bias collected. 
The physics reach is defined as the
relative cross-section of a process (with respect to the minimum bias one)
times the global efficiency for LHCb to detect it.
As it can be seen $K^0_S$ and $\Lambda$ production can start to be
studied  with about $10^6$ minimum bias events (\textit{i.e.} seconds of data
taking); this study is described in §\ref{sec:v0}.
With $10^7$ minimum bias $D$ production can be studied, as described in
§\ref{sec:D}, as well as $J\psi$ production (§\ref{sec:jpsi}), while for $b\to
J/\psi$ something more than $10^8$ minimum bias events will be needed.
This last sample, $10^8$ events, is taken as a reference for the performances 
of the analysis studies presented in the following. It is to be bared in mind
that such a sample corresponds to days of data taking at nominal conditions, so
these studies will be performed as soon as the proton collisions will take
place; in particular the mentioned physics channels, apart from being
interesting by themselves, are propaedeutic
for the main analyses of LHCb, helping either to understand the detector 
performances and as building blocks of $b$-hadrons decays (\emph{e.g.} $B\to
J\psi
K^0_S$).

\section{V0 production \label{sec:v0}}

The hadronization process is still not well understood within the
theoretical frame. 
At hadron colliders the strange quark production can give hints on the
hadronization process since strange quarks are not present in the initial state
valence quarks.
Various phenomenological models are available, which are often
different as philosophy, that can reproduce the fragmentation into strange
hadrons and explain phenomena and distributions. 
Within Monte Carlo event simulators (e.g. PYTHIA) different tunings are
available; but, although they do agree at Tevatron energies, there are
divergences in
the extrapolation at the LHC energies.
Hence the study of V0 production at LHC (namely $\Lambda$'s and $K^0_S$) can
shed some light in these processes.  

There are also experimental motivations that justify V0 production studies.
First of all these decays need minimal requirements from
the detector point of view. Just information from the VELO and the
Tracker is needed; moreover the simple minimum bias trigger will be
available from the beginning of data taking.
Being $\Lambda$'s and $K^0_S$ masses well known,  
crosschecks of the momentum calibration can be made against invariant masses.
No particle identification is required so that from these samples RICH
calibration will be available as well. 
Finally these particles are building blocks of subsequent and more complex
analysis.

\subsection{V0 analysis \label{sec:v0analysis}}

The study of V0 particles in LHCb will start from the following channels: 
$K^0_S \to \pi^{+} \pi^{-}$, $\Lambda \to p \pi^{-}$,
$\bar \Lambda \to \bar p \pi^{+}$.
The results that will be presented in the following come from a study based
on simulated minimum bias events with full detector simulation.

In order to deal with clean events, the ones with just one primary vertex
have been selected.
The analysis starts from combining two opposite charged tracks, in particular
just \emph{long} tracks are used: \textit{i.e.} tracks with hits along
the whole Tracker and in the VELO.
As already said, no particle identification information is required, the
analysis being based only on geometry and kinematics. 

\begin{figure}
\includegraphics[width = 8 cm]{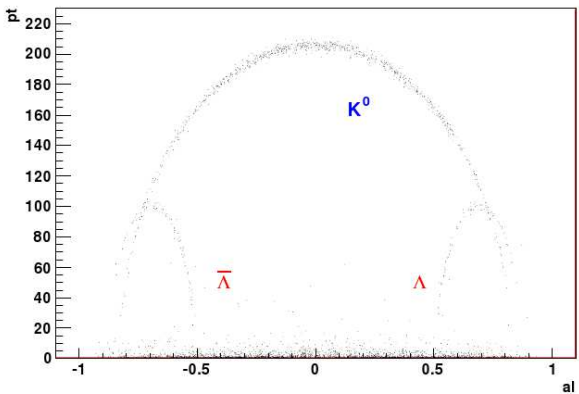}
\caption{Armenteros-Podolanski plot for minimum bias events selected as
described in §\ref{sec:v0analysis}. Typical $K^0_S$, $\Lambda$ and $\bar
\Lambda$ ellipses can be seen, apart from some combinatorial background in the
lower part of the plot.}\label{fig:armenteros}
\end{figure}

The $K^0_S$ analysis requires just the following two simple cuts: one on the
distance of closest approach (DOCA) between the two tracks (required to be less
than 0.2 mm) and one on the $K^0_S$ proper time $c\tau>4~\mr{mm}$.
The $\Lambda$ selection imposes the same cuts, plus a cut on the impact
parameter of the $\Lambda$ with respect to the Primary Vertex (PV), which is
required to be less than 0.1 mm. 
In order to distinguish between $K^0_S$, $\Lambda$ and $\bar \Lambda$ decays,
without particle identification, the Armenteros-Podolanski plot will be used: 
the transverse momentum ($p_T$) of decay products with respect to the
mother
particle is plotted versus the longitudinal momentum asymmetry, which is
defined as 
$$
\alpha = \frac{p^{+}_L - p^{-}_L}{p^{+}_L + p^{-}_L}
$$ 
where $p^{\pm}_L$ is the longitudinal momentum (with respect to the mother's
direction) of the daughters particles. The Armenteros-Podolanski plot, for the
studied samples, is shown in Fig.~\ref{fig:armenteros} where the three signal
regions can be easily distinguished one from each other
and from the combinatorial background in the lower part of the plot.
 Giving the possibility to separate the
three mentioned decays, this technique leads to kinematic particle
identification so that unbiased pions and protons samples are available for PID
(and hence RICH) calibration.

\begin{figure}
\subfigure[]{\includegraphics[width = 8 cm]{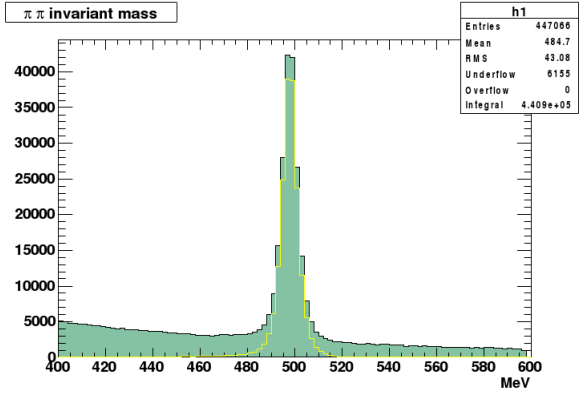}}
\subfigure[]{\includegraphics[width = 8 cm]{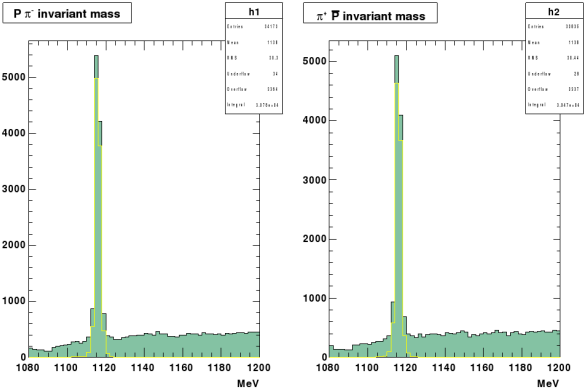}}
\caption{Invariant mass distributions for $K^0_S$ (a) and
$\Lambda$ (b) decays after the selection described in
§\ref{sec:v0analysis}.}\label{fig:v0masses}
\end{figure}

After this very simple analysis, the invariant mass distributions for $K_S$ and
$\Lambda$'s can show already very prominent peaks with good signal to background
ratios; the distributions are shown in Fig.~\ref{fig:v0masses}.
The production measurement of V0 particles will be made in $p_T$ and
pseudorapidity ($\eta$) bins. Background subtraction will be calculated for
each bin using the invariant mass distribution sidebands fit. 
Within a typical sample of $10^8$ minimum bias events
around $2\cdot 10^6$ reconstructed $K^0_S$ are expected and around $10^5$
$\Lambda$ and $\bar \Lambda$.

\begin{figure}
\subfigure[$\bar \Lambda/ \Lambda$ ratio for different
``Perugia'' tunings]{\includegraphics[width
= 8 cm]{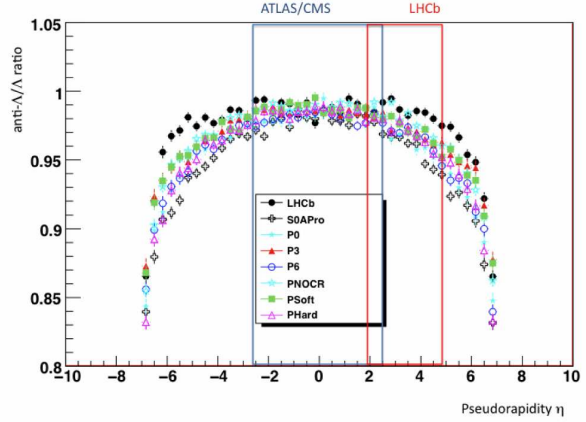}}
\subfigure[$\bar \Lambda/ \Lambda$ ratio relative error]{\includegraphics[width
= 8
cm]{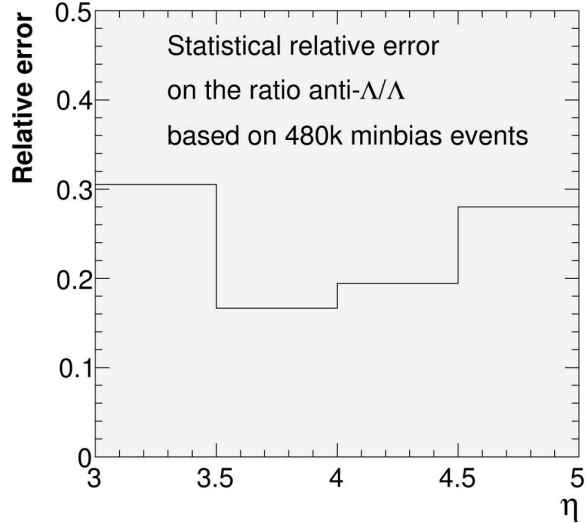}}
\caption{Barionic asymmetry: (a) $\bar \Lambda/ \Lambda$ ratio as obtained from
 different tunings of the Monte Carlo generator; it can be seen that the LHCb
coverage is in a more sensitive region in respect to ATLAS and CMS; (b) Plot of
the relative error on the $\bar \Lambda/ \Lambda$ ratio, as a function of the
pseudorapidity, as obtained using 480k simulated minimum bias events.}
\label{fig:barionic}
\end{figure}

Apart from production studies, also the baryonic asymmetry
will be measurable in the $\Lambda$ sector. 
The $\bar \Lambda/\Lambda$ ratio, in fact, provides
discrimination between different hadronization processes \cite{ref:skands}. In
Figure~\ref{fig:barionic} (a) the
theoretical expectation for $\bar \Lambda/\Lambda$ as a function of $\eta$ is
shown as calculated with different tunings of the Monte Carlo generators. As it
can be seen from the plot, the LHCb experiment will have
a larger sensitivity with respect to other LHC experiments due to its coverage
in the forward region of $p$-$p$ interactions.
A first analysis within the LHCb experiment, based just on 480k Monte Carlo
minimum bias events leads to the plot shown in Figure~\ref{fig:barionic}
(b): the
relative error on the determination of $\bar \Lambda/\Lambda$ ratio is shown as
a function of the pseudorapidity. As it can be seen the error is already at the
level of 20-30\% which, extrapolated to the reference sample of $10^8$
minimum
bias events, leads to expected statistical errors at the level of 1.5\% which
will provide discrimination between different models. 

\section{D meson production studies \label{sec:D}}

\begin{figure}
 \subfigure[D0 invariant mass: cut-based analysis]{\includegraphics[width = 7
cm]
{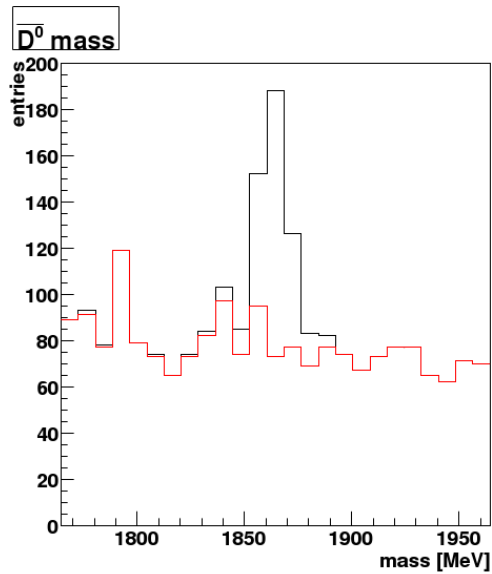}}

\subfigure[D0 invariant mass: MVA method]{\includegraphics[width = 7
cm]
{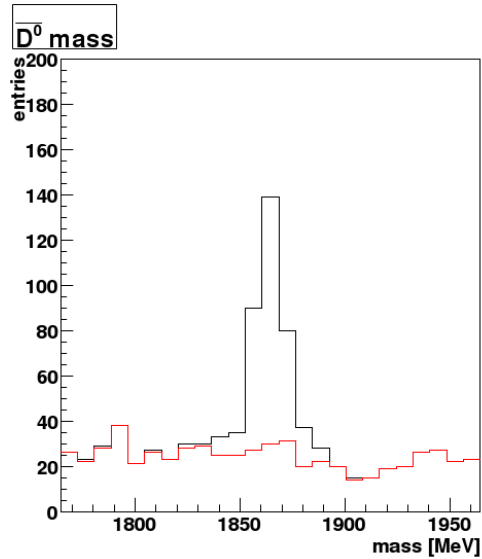}}
\caption{D0 invariant masses distributions as obtained from a standard
cut-based analysis (a) and from a Multivariate analysis
(b). In both, in red is shown the background distribution
and in black the sum of signal and background.}\label{fig:d0masses}
\end{figure}

During the first days of data taking, as well as $strange$ production, $charm$
production will also be studied.
In particular early studies on D mesons production will be made, by means of
the following decays: $D^0 \to K^{-} \pi^{+}$ and $D^{+} \to K^{-} \pi^{+}
\pi^{+}$ plus charge conjugate modes.
The analysis strategy for the selection of these decays is very similar to the
one for V0 searches. Again only VELO and Tracker information are required, no
particle identification being needed from RICH, and using only kinematic and
geometric variables. No use of \emph{significance} variables (\textit{i.e.}
variables divided by their errors) will be made in order to avoid biases coming
from errors not properly understood during the early stage of the experiment.
Moreover particle ratios will only be
studied in order to cancel out largely systematics.\\
Finally a use of Multivariate analysis techniques is made in order to reduce
background. The particular algorithm used in this case is called
\textsc{Ripper}, which is a rule based classifier.
Analysis studies have been based on 9.5 M Monte Carlo minimum
bias events. Taking the $\bar D^0 \to K^+ \pi^-$ decay as an example,
the
following variables have been considered both for a traditional cut based
analysis and for a Multivariate analysis: $p_T$ of the daughter
particles, $p_T$ of the $D^0$, impact parameters of daughters
with respect to PV and angle between the two IP vectors, $D^0$ flight
length, DOCA, impact parameters of $D^0$. The results of the analysis
can be seen in Fig.~\ref{fig:d0masses}, where the invariant mass distributions
obtained by selecting events with the two techniques are shown: by setting the
Multivariate method to have the same efficiency on the signal as the cut based
method, the background is reduced by the former of a
factor~$\sim 3$. 

From the analyzed sample an average of about 200 events for each species has
been, for the expected analysis sample of $10^8$ minimum bias events
around 2000 events for each species are foreseen. 
This results lead to an expected error on the particle ratios in the
sensitivity region of $\sigma(\frac{\bar D^{0}}{D^{0}}) = 5\% $ and
$\sigma(\frac{D^{-}}{D^{+}}) = 6\%$.

\section{$J/\psi$ production \label{sec:jpsi}}

The production of prompt $J/\psi$ is not completely understood. 
Using the Non-Relativistic QCD (NRQCD) with Colour Octet Model is possible to
reproduce the transverse momentum spectrum measured at Tevatron; unfortunately
this model foresees also an increasing of the transverse polarization with the
$p_T$, which has not been seen at Tevatron. Alternative models (\textit{e.g.}
Colour Evaporation) also predict the same $p_T$ spectrum but cannot account for
the polarization. Hence cross-section and polarization measurements are
important to understand charmonium production. Given the big cross-section of
$J/\psi$ production and the unique LHCb
coverage, large data samples will be available already in the first days of LHC
running. At the same time also the production of $J/\psi$ from $b\bar b$ decays
will be
measured. 

\subsection{$J/\psi$ selection \label{sec:jpsisel}}

The easiest $J/\psi$ decay to look for in order to measure its production is
the one into two muons. Even if, as already said, many $J/\psi$ will be already
present in the minimum bias sample, the use of
the Muon trigger, which is one of the main blocks of the LHCb trigger,
will allow to have many more events. 
The selection of $J/\psi$ candidates starts from two opposite \emph{long}
tracks with hits also in the Muon stations. 
A cut is made on the $\chi^2$ of the track and of the vertex of the two tracks. 
One of the two muons is required to have a transverse momentum greater than 1.5
GeV. Moreover a cut on the identification likelihood ($L$) is applied, with
an efficiency on the muons $\varepsilon_{\mu} \sim 90 \%$ at a level of pion 
mis-identification of $\varepsilon_{\pi} \sim 1.4\%$.

\begin{figure}
 \includegraphics[width  = 8 cm]{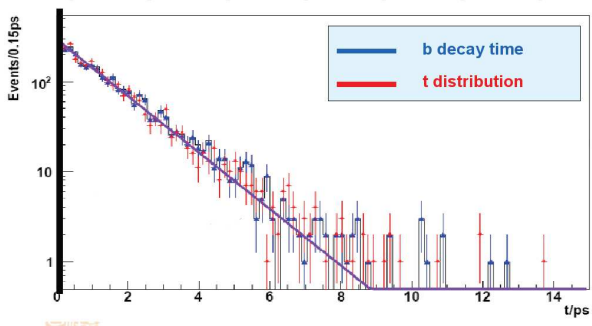}
\caption{$b$ quark lifetime distribution studied at the event generator level
in blue and $t$ variable distribution in red; it can be seen that the two
distribution are very similar. Prompt $J\psi$ events in this plot would lie on
the very far left.}
\label{fig:blifetime}
\end{figure}

\begin{figure}
\includegraphics[width = 8 cm]{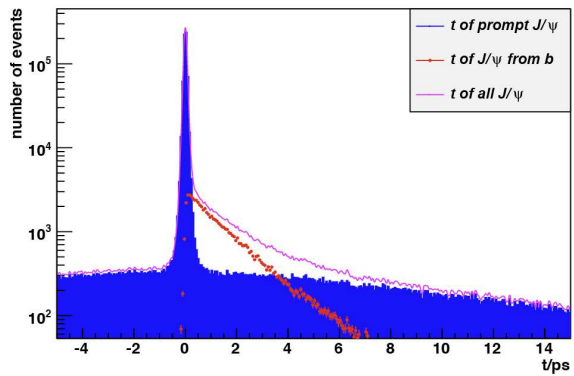}
\caption{Distribution of the $t$ variable for
different components: the blue curve is the prompt
$J\/psi$ distribution, the red curve are $J\psi$ from $b$, and the magenta the
sum of the two. Within the blue curve, the long tail is due to the combination
of true $J\psi$ with wrong Primary Vertices.}\label{fig:tdistribution}
\end{figure}

About 8\% of the $J/\psi$ produced at LHC are expected to come from
$b$-hadron decays; in order to distinguish between these and the prompt
$J/\psi$'s 
the following variable will be used: 
$$
t  = \frac{z_{J/\psi}- z_{PV} }{p_z^{J/\psi}}\cdot m_{J/\psi} 
$$
where $z_{PV}$ and $z_{J/\psi}$ are the two vertices coordinates and 
$p_z^{J/\psi}$ the momentum both along $z$ and $m_{J/\psi}$ the mass of the
$J/\psi$. From studies at the generator level $t$ is known to be a good
approximation of
the $b$-quark lifetime. In particular in
Fig.~\ref{fig:blifetime} the
distribution of $t$ is shown together with the $b$-decay time and the two
distributions are close one to each other.
In Fig.~\ref{fig:tdistribution} instead, the distribution of the $t$
variable
for different  $J/\psi$ contributions is shown. In particular, apart from a
flat combinatorial background (the shape of which will be estimated from
$J/\psi$ invariant mass sidebands), the prompt $J/\psi$ peak at $t=0$ and the 
 $J/\psi$ from $b$ exponential tail can be distinguished. 
Within the long tail of the distribution there is also a component due to 
true prompt $J/\psi$ associated to the wrong primary vertex giving so a wrong
$t$
value. The amount of this component will be estimated by using
true reconstructed $J/\psi$ and associating them with primary vertices coming
from other events chosen randomly. 


\subsection{$J/\psi$ measurement}

\begin{figure}
\includegraphics[width = 8 cm ]{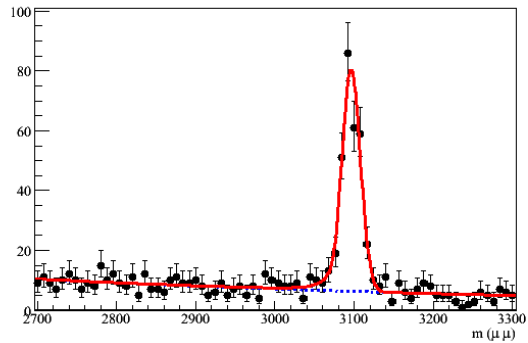}
 \caption{$J/\psi$ invariant mass distribution (in MeV) as obtained
by applying the selection sketched in §\ref{sec:jpsisel} to
a sample of 19 millions minimum bias events.}\label{fig:jpsimass}
\end{figure}

Using the first $5 ~\mr{pb}^{-1}$ of integrated luminosity 3 million
reconstructed
$J/\psi$ are expected. The event counting will be made in pseudorapidity and
transverse momentum bins.
In order to compute the number of $J/\psi$ in each bin a fit of the invariant
mass distribution will be made. 
In particular the plot shown in Fig.~\ref{fig:jpsimass} shows this distribution
obtained by just using 19 millions minimum bias events (corresponding to seconds
of data taking at nominal luminosity), the mass resolution obtained is about 11
MeV.
Monte Carlo data will be used to correct for detector acceptance and
efficiencies for trigger and offline selection.
Finally in order to measure the absolute cross-section, integrated luminosity
measurements will be needed.

As already said, an important parameter to measure will be the 
$J/\psi$ polarization $\theta$; this is defined as the angle between the
$\mu^+$ momentum in the $J/\psi$ rest frame and the $J/\psi$ momentum in the
lab frame.
In the helicity frame the distribution is parametrized as 
$$
\frac{dN}{d
\cos(\theta)} \propto 1+\alpha \cos^2 (\theta)
$$ 
where $\alpha = 1$ means transverse polarization, $\alpha = -1$ longitudinal
polarization and $\alpha = 0$ no polarization.
Simulation studies have shown that the LHCb geometry and L0 Trigger efficiency
will induce a fake polarization on the $J/\psi$ reconstructed sample, even if
the original sample were not polarized; so that in order to measure the $J/\psi$
cross-section, its polarization should be taken into account.
Polarization measurement then will be made and will be also used to discriminate
between different production models. 

\section{$\chi_c$ production \label{sec:chi}}

From Tevatron measurements it's known that about $30 \%$ of $J/\psi$ come
from $\chi_{c(1,2)} \to J/\psi \gamma$ and hence have different polarization.
The production of $\chi_{c(1,2)}$ is also interesting by itself being the ratio 
$R_{\chi_c} = \sigma(\chi_1)/\sigma(\chi_2)$ important to distinguish different
production models. While the NRQCD predicts $R_{\chi_c}$ to be close to 1,
Colour Evaporation Model predicts it to be $5/3$ from the
expectation from spin-counting.

\begin{figure}
\subfigure{\includegraphics[width = 8 cm
]{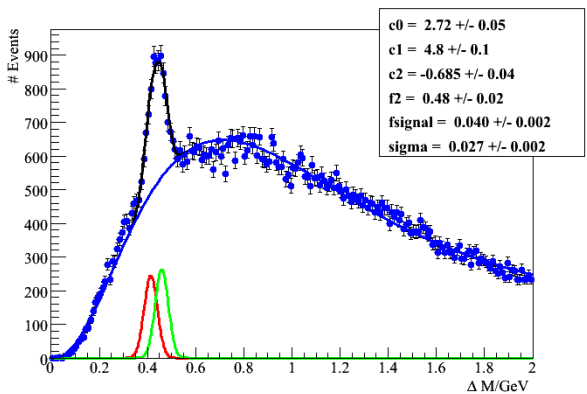}}
\caption{Distribution of the $\Delta M$, as defined in §\ref{sec:chi},
for inclusive $J/\psi$ events. The $\Delta M$ for $\chi_{c(1,2)} \to J/\psi
\gamma$ events peak can be seen. The red and green curves are the gaussian
distribution used to represent the signal; the blue curve represents the
background ($J/\psi$ not from $\chi_c$) distribution.}\label{fig:chideltam}
\end{figure}

In order to select $\chi_{c}$ candidates a photon with $p_T > 500$ MeV is
combined with a $J/\psi$ selected as before (restricted to a $\pm 40$ MeV mass
window).
The distribution of $\Delta M = M_{\mu \mu \gamma} - M_{\mu \mu}$ is then
used to extract the $\chi_{c1}$ and $\chi_{c2}$ signals.
This distribution, shown in Fig.~\ref{fig:chideltam}, will be fitted
with two
gaussians which represent the signal and a background component parametrized as 
$P(\Delta M) = \Delta M^{c_0} \cdot \exp (- c_1 \cdot \Delta M - c_2 \cdot 
\Delta M^2 )$. 
The obtained resolution is $\sigma (\Delta M) \sim 27$ MeV which has to be
compared with the mass difference of the two resonances, $m(\chi_{c2}) -
m(\chi_{c1}) = 46$ MeV; the achievable separation is limited but still some
sensitivity is present leading to a discrimination between  $\chi_{c1}$ and
$\chi_{c2}$ and to the measurement of $R_{\chi_c}$. 

\section{Other quarkonia measurements}

Beyond the already mentioned studies, measurements will be performed also on
other charmonia and bottomonia channels. The $\psi(2S)$ production and in
particular the $\sigma(\psi(2S))/\sigma(J/\psi)$ ratio will be studied. 
Moreover, similar measurements to the charmonia case will be performed
also for bottomonium production and spectroscopy (\textit{e.g.} 
$\varUpsilon (1S,2S,3P)$ resonances, $\chi_b \to \varUpsilon \gamma$).

The exotic $X$, $Y$ and $Z$ charmonia states, recently observed, will also be
studied at LHCb. In particular is under investigation the possibility to
measure the $X(3872)$ quantum numbers by means of an angular analysis of the
decay $X(3872) \to J/\psi \pi^+ \pi^-$; this analysis will be even more
sensitive if studied in the frame of the $B^+ \to X K^+ $ decay where the $X$ is
expected to be produced polarized: in this context will be possible to
distinguish if the $X$ has $J^{PC}$ equal to $1^{++}$ or $2^{-+}$.

\section{Conclusions}

To summarize, the LHCb experiment, ready to take data at the LHC proton
collisions, will be able to exploit the very first data, apart for
detector calibrations and tunings, also for very interesting physics analyses.
Within a sample of $10^8$ minimum bias events it will be possible to probe
hadronization models with V0 production studies and D-mesons ratios will be
measurable with about $\sim 5\%$ error. Quarkonia
spectroscopy will be also investigated and $J/\psi$ production and polarization
will be measured.


\bigskip 

\end{document}